
\magnification=1200

\baselineskip=18pt 

\nopagenumbers
\voffset=1truecm
\hsize= 5.25in 
\vsize=7.5in 

\def\({\c c }

\def\|{\'\i}

\def\nl{\par\noindent}

\pretolerance=10000

\centerline{\bf PHASE TRANSITION IN  SCALAR THEORY }
\centerline {QUANTIZED ON THE LIGHT-FRONT$^{a}$}

\vskip 0.4cm
\centerline {Prem P. Srivastava\footnote{*}{
E-mail:  prem@cbpfsu1.cat.cbpf.br\qquad\qquad {\sl HADRONS94, Gramado, RS,
1994}
\nl {\sl \qquad To be published in: {\it Proceedings of the Workshop on Hadron
Physics, 1994}, Gramado, RS, World Scientific, Singapore. }  }}
\vskip 0.2cm                                      
\centerline {\sl Instituto de F\|sica, Universidade do Estado do
Rio de Janeiro, RJ}

\vskip 0.4cm

\nl {\bf Abstract:}\quad {\sl The renormalization of the two dimensional
light-front quantized  $\phi^{4}$ theory is discussed.
The mass renormalization condition and the renormalized
constraint equation are shown to contain all the information
to describe the phase transition in the theory, which is found to be
of the second order. We argue that the same
result would also be obtained in the conventional equal-time formulation.
  }

\vskip 5truecm 
\vfill
\nl IF-UERJ-0016/94

\eject

\pageno=1

\baselineskip=18pt
\bigskip

\nl {\bf 1.} We use the light-front framework [1] to
study
the  stability of the vacuum in the $\phi^{4}$ theory
when the coupling constant is increased from vanishingly small values to
larger values. The hamiltonian formulation now has an additional
ingredient in the form of a constraint eq. It is not convenient to
eliminate the constraint; it would require to handle a nonlocal hamiltonian
to construct renormalized theory. We could alternatively perform the
ususal renormalization and obtain instead a {\it renormalized constraint eq.}.
This along with, say, the mass renormalization condition may furnish important
information on the nature of the phase transition in the theory.
We recall that there are rigorous proofs on the triviality of
$\phi^{4}$ theory in the continuum for more than four space-time
dimensions and on its interactive nature for dimensions less than four.
In the important case of four dimensions the
situation is still unclear   and light-front dynamics
may throw some light on it. In view of the
complexity of the renormalization problem in this case
we will illustrate our points by
considering  only the two dimensional theory, which is of importance
in the condensed matter physics.  For example, from the well established
results on the generalized Ising models, Simon and Griffiths [2]
conjectured some time ago that the two dimensional $\phi^{4}$ theory should
show the  {\it second order} phase transition. We do confirm it here perhaps
for the first time.  The earlier works [3] based on variational
methods usually in the conventional equal-time formulation, viz.,
the Hartree approximation, Gaussian effective potential,
using a  scheme of canonical transformations etc.,
give  rise to the first order
phase transition contradicting the conjecture.

\bigskip
\noindent {\bf 2.} We renormalize the theory based on
(3-1) and (4-1) of ref. [1] without solving the
constraint eq. (4-1) first to eliminate $\omega$.
We set   $M_{0}^{2}(\omega)= (m_{0}^{2}+3 \lambda \omega^2)$
and choose   ${\cal H}_0={M_0}^2\varphi^2/2$
so that  ${\cal H}_{int}= \lambda\omega\varphi^3+\lambda
\varphi^4/4$. In view of the  superrenormalizability of the two dimensional
theory we need to do only  the mass renormalization.
We assume that the bare mass $m_{o}$  is nonvanishing and set
$\int dx \varphi(x,\tau)={\sqrt{2\pi}}\,\tilde\varphi(k=0,\tau)=0$, e.g.,
$k\equiv {k^{+}}>0$.
We  follow [3] the  straightforward
Dyson-Wick expansion  based on the Wick theorem in place of the old fashioned
perturbation theory.

The  self-energy correction  to the  {\it one loop order} is
$-i\Sigma(p)=-i\Sigma_1-i\Sigma_2(p)
=(-i6\lambda){1\over 2} D_{1}({M_0}^2)+(-i6\lambda\omega)^2 {1\over 2}
(-i) D_2(p^2,{M_0}^2)\,$,
where the divergent contribution  $D_1$
refers to the one-loop tadpole while  $D_{2}$ to the
one-loop  {\it finite contribution} coming from the $\varphi^{3}$ vertex.
The latter carries the sign {\it opposite}
to that of the first and it will be argued
below to be of the same order in $\lambda$
as the first one. Using the dimensional regularization  we obtain

$$ {M_0}^{2}(\omega)=M^{2}(\omega)+{{3\lambda}\over
{4\pi}}\Bigl[\gamma+ln({M^2(\omega)\over {4\pi\mu^2}})
\Bigr]+18\lambda^2\omega^2 D_2(p,M^2)\vert_{p^2=-M^2}
+{3\lambda\over {2\pi}}{1\over {(n-2)}}.\,\eqno(1)$$

\noindent Here $M(\omega)$ is the physical mass and
we  take into account that in view of the tree level
result
$\,\omega(\lambda\omega^2+{m_0}^2)=\omega[{M_0}^2(\omega)-2\lambda\omega^2]
=0$ the correction term  $\lambda^2\omega^2 $, for  $\omega \ne 0$,  is
really  of the first order in
$\lambda$. We ignore terms of order $\lambda^{2}$ and higher
and  remind  that
$M_{0}$ depends on $\omega$ which in its turn is involved in the
constraint eq.
{}From  (1) we obtain the {\it mass renormalization condition }

$$M^2-m^2=3\lambda\omega^2+{{3\lambda}\over {4\pi}}
ln({m^2\over {M^2}})-\lambda^2
\omega^2 {\sqrt{3}\over {M^2}}\,\eqno(2)$$

\noindent where $M(\omega)\equiv M$ and
$M(\omega=0)\equiv m$ indicate the phsyical masses in the
{\it asymmetric} and {\it symmetric} phases respectively.

We next take the vacuum expectation value of the
constraint eq. (4-1)  in order to obtain another independent equation.
To the lowest order  we find
{\it renormalized constraint eq.}

$$\beta(\omega)\equiv\omega\Bigl[M^2-2\lambda\omega^2 +
\lambda^2\omega^2 {{\sqrt 3}\over M^2}
-{{6\lambda^2}\over (4\pi)^2}{b\over M^2}\Bigr]=0.\,\eqno(3)$$

\nl where $b$ arises from $D_3$,
a finite integral like $D_{2}$ with three denominators, and
$b\simeq 7/3$.

The expression for
{\it difference}  of the renormalized {\it vacuum energy density},
in the equal-time formulation,
in the broken  and the symmetric phases, is found to be finite and given by

$$\eqalignno{F(\omega)&={\cal E}(\omega)-{\cal E}(\omega=0)\cr
&={(M^2-m^2)\over {8\pi}}+{1\over {8\pi}}(m^2+3\lambda\omega^2)
\,ln({m^2\over M^2})+{3\lambda\over 4}\Bigl[{1\over {4\pi}}ln({m^2\over M^2})
\Bigr]^2\,\cr
&\qquad\quad+{1\over 2}m^2 \omega^2+{\lambda\over 4}\omega^4+{1\over {2!}}.
(-i6\lambda\omega)^2.{1\over 6}.D_3(M),\quad&(4)\cr} $$

\nl  which is also
independent of the arbitrary mass $\mu $ when we use  the mass renormalization
condition.
We verify  that $(dF/d\omega)=\beta$
and $d^2F/d\omega^{2}=\beta'$ and except for the finite last term
it coincides with the result in the earlier works.
The last term in $\beta$
corresponds to a correction $\simeq \lambda(\lambda\omega^2)$ in this energy
difference and  may not be ignored  like in the case of the
self-energy.
In the equal-time case   (3) would be
required to be {\it added} as an external constraint to the theory based
upon physical considerations. It will  ensure that
the sum of the tadpole diagrams, to the approximation concerned,
for the
transition $\varphi\to vacuum$ vanishes. {\it The physical
outcome would then be
the same in the two frameworks of treating the problem  considered}.
The variational methods write only the first two ($\approx$ tree level)
terms in the  expression for $\beta$  and thus ignore
the terms coming from the finite corrections.
A similar remark can be made about the last term in (2).

Consider first the {\it symmetric phase} with $\omega\approx 0$, which is
allowed by (3).
{}From (2) we  compute
$\partial M^2/\partial{\omega}=
2\lambda\omega(3-{\sqrt 3}\lambda/M^2)/[1+3\lambda/(4\pi M^2)-{\sqrt 3}
\lambda^2\omega^2/M^4]$ which is needed to find  $\beta'\equiv
d\beta/d\omega =d^2F/d\omega^{2}$.
Its  sign will determine the nature of the stability
of the vacuum.  We find
$\beta'(\omega=0)=M^2[1-0.0886(\lambda/M^2)^2]$, where by
the same arguments as made above in the case of $\beta$ we may not ignore
the $\lambda^{2}$ term. The $\beta'$
changes the sign from a positive value for vanishingly weak couplings to
a negative one  when the coupling increases.  In other words
the system starts out in a {\it stable symmetric phase}
for vanishingly    small coupling but
passes over into an {\it unstable symmetric phase} for values greater
than $g_{s}\equiv\lambda_{s}/(2\pi m^2)\simeq 0.5346$.

Consider next the case of {\it the spontaneously broken symmetry
phase} ($\omega\ne 0$). From (3) the values of $\omega$ are now given by
$M^2-2\lambda\omega^2
+{\sqrt 3}\lambda/ 2=0,\,$
where  we have used the tree level approximation
$\, 2\lambda \omega^2\simeq M^2\,$.
The mass renormalization condition becomes
$M^2-m^2=3\lambda\omega^2+({3\lambda}/({4\pi}))\,
ln({m^2/ {M^2}})-\lambda {\sqrt{3}/2}$.
On eliminating $\omega$
we obtain the {\it modified duality relation}

$${1\over {2}}M^2+m^2+{{3\lambda}\over {4\pi}}\,
ln({{m^2}\over {M^2}})+{{\sqrt 3}\over 4}\lambda=0.\,\eqno(5)$$

\noindent which can also be rewritten as
$[\lambda\omega^2+m^2+(3\lambda/({4\pi}))
ln(m^2/M^2)]=0$ and it shows that  the  real solutions exist only for
$M^2 > m^2$. The finite corrections found here are again
not considered in the earlier works, for
example, they  assume (or find)  the tree level expression
$\,M^{2}-2\lambda\omega^{2}=0$.
In terms of  the dimensionless coupling constants
$g=\lambda/(2\pi m^2)\ge 0$ and $G=\lambda/(2\pi M^2)\ge 0$
we  have  $G<g$.
The new self-duality eq.  differs from the old one  and
shifts  the critical coupling to a higher value.
We find that: {\it i)} for $g < g_c=6.1897$ ({\it critical coupling})
there is no real solution for $G$, {\it ii)} for a fixed $g>g_{c}$ we have two
solutions for $G$ one with the point lying on the upper branch
($G>1/3$) and the other with that on the lower branch ($
G<1/3$),  of the curve describing $G$ as a function of $g$ and
 which starts at the point $(g=g_{c}=6.1897,G=1/3)$,
{\it iii)} the lower branch with $G<1/3$,
approaches to a vanishing value for $G$ as $g\to\infty$, in contrast to the
upper one for which $1/3<G<g$ and  $G$ continues to increase.
{}From  $\beta(\omega)\approx \omega[M^2-2\lambda\omega^2
+{{\sqrt 3}\lambda/2}]\,$ and (5) we  determine
$\beta'\approx (1+0.9405 G)$ which is always  positive and thus
indicates a minimum for the difference $F(\omega)$
of the vacuum energy densities (
when $\omega$ is nonvanishing).
The energetically favored broken symmetry phases
become available only after the
coupling grows to the critical coupling $g_c=6.18969$
and beyond this the asymmetric phases would be preferred against
the unstable symmetric phase in which the system finds itself when
$g>g_{s}\simeq 0.5346$. The phase transition is thus of the {\it second order}
confirming the conjecture of Simon-Griffiths. If we ignore the additional
finite renormalization corrections  we obtain
complete agreement with the earlier results and a first order phase
transition, e.g., the symmetric phase always remains stable but for
$g>1.4397 $ the energetically favoured symmetric phases also do appear.
{}From numerical computation we verify that at the minima
corresponding to the nonvanishing value of $\omega$
the value of F is negative
and that for a fixed $g$ it is more negative for the point on the
lower branch ($G<1/3$) than for that on the upper branch
($G>1/3$).
\bigskip

\nl {\bf 3.} The  present work and the earlier one
on the mechanism of {\it SSB}  add
to the previous experience that the {\it front form} dynamics is a useful
complementary method and needs to be studied systematically in the context of
QCD and other problems. The physical results following from
one or the other form of the theory
should come out to be the same though the mechanisms to arrive at them
may be  different. In the equal-time case
we are required to add external considerations in order
to constrain the theory
while  the analogous conditions in the light-front formulation
seem to be already contained in it through the  self-consistency equations.
When additional fields, e.g., fermionic ones are present also
the constraint equations in the theory quantized on the light-front
would relate the various types of vacuum condensates (vacuum expectation
values of composite scalar fields).
\bigskip

\nl {\bf References}:

\item{a.} Shortened from   Nuovo Cimento A, 1994;
also one submitted to Physics Letters B.

\item{[1.]} The Hamiltonian (3-1) and the constraint eq. (4-1) used here
 derived  in the previous contribuition (hep-th@xxx.lanl.gov/9412204)
{\it Light-front quantization and
Spontaneous symmetry breaking} are

$${ P^{-}\,=\int  dx \,\Bigl [\omega(\lambda\omega^2-m^2)\varphi+
{1\over 2}(3\lambda\omega^2-m^2)\varphi^2+
\lambda\omega\varphi^3+{\lambda\over 4}\varphi^4
\Bigr ]\,}\eqno(3)$$

$$ \eqalign {& \,lim_{L\to\infty}
{1\over L}\int_{L/2}^{L/2} dx \,V'(\phi)\equiv  \cr
& \omega(\lambda\omega^2-m^2)+lim_{L\to\infty} {1\over L}
 \int_{-L/2}^{L/2} dx \Bigl[ \,(3\lambda\omega^2-m^2)\varphi +
 \lambda (3\omega\varphi^2+\varphi^3 ) \,
\Bigr]=0}\,\eqno(4)$$

\item{[2.]} B. Simon and R.B. Griffiths, Commun. Math. Phys. {\bf
33} (1973) 145;  B. Simon, {\it The ${P(\Phi)_{2}}$ Euclidean (Quantum)
Field Theory}, Princeton University Press, 1974.
\item{[3.]} See for list of
references: {\it Lectures on
 Light-front quantized field theory}, Proceedings {\it XIV
 Brazilian National Meeting on Particles and Fields},
 Sociedade Barasileira de F\|sica, pgs. 154-192, 1993.  Available
also from hep-th@xxx.lanl.gov, no.  9312064.

\bye